\input{aipcheck}

\documentclass[
    ,final            % use final for the camera ready runs
  ]
  {aipproc}

\layoutstyle{6x9}

\begin{document}

\title{Nuclear shell structure and response toward the limits of mass, temperature and isospin
}



\author{E. Litvinova}{
  address={National Superconducting Cyclotron Laboratory, Michigan State University, East Lansing, MI 48824-1321, USA},
%  address={Physics Department, Western Michigan University, Kalamazoo, MI 49008-5252, USA},
%  altaddress={National Superconducting Cyclotron Laboratory, Michigan State University, East Lansing, MI 48824-1321, USA}
   }

\author{B.A. Brown}{
  address={Department of Physics and Astronomy and National Superconducting Cyclotron Laboratory, Michigan State University, East Lansing, MI 48824-1321, USA}
}

\author{D.-L. Fang}{
  address={National Superconducting Cyclotron Laboratory, Michigan State University, East Lansing, MI 48824-1321,
  USA}, altaddress={Joint
Institute for Nuclear Astrophysics, Michigan State University, East
Lansing, MI 48824-1321, USA} }

\author{T. Marketin}{
  address={Physics Department, Faculty of Science, University of Zagreb, 10000 Zagreb, Croatia}
}

\author{P. Ring}{
  address={Physik-Department der Technischen Universit\"{a}t
M\"{u}nchen, D-85748 Garching, Germany}}

\author{V.I.~Tselyaev}{
  address={Nuclear Physics Department, St. Petersburg State
University, 198504 St. Petersburg, Russia} }

\author{R.G.T. Zegers}{
  address={Department of Physics and Astronomy and National Superconducting Cyclotron Laboratory, Michigan State University, East Lansing, MI 48824-1321,
  USA},
 altaddress={Joint
Institute for Nuclear Astrophysics, Michigan State University, East
Lansing, MI 48824-1321, USA}}

\begin{abstract}
We present a short overview of our recent theoretical developments
aiming at the description of exotic nuclear phenomena to be reached
and studied at the next-generation radioactive beam facilities.
Applications to nuclear shell structure and response of nuclei at
the limits of their existence, with a special focus on the physics
cases of astrophysical importance, are discussed.
\end{abstract}

\maketitle

%%%%%%%%%%%%%%%%%%%%%%%%%%%%%%%%%%%%%%%%%%%%
%% MAINMATTER
%%%%%%%%%%%%%%%%%%%%%%%%%%%%%%%%%%%%%%%%%%%%

%\section{<A section>}
%
Last decades, low-energy nuclear physics has expanded considerably
its domain due to the opportunities opened by rare isotope beam
facilities of the new generation. In particular, the techniques of
the isotope separation online and in-flight production have been
implemented at the major low-energy nuclear physics facilities with
great success: numerous experiments on synthesis of exotic nuclei
and on studies of their dynamical properties have been performed.
This has produced a strong catalyzing effect on theoretical
developments toward finding a high-precision and highly universal
solution of the nuclear many-body problem. In this contribution, we
give a brief overview of our recent developments and applications to
nuclear systems toward the limits of mass, isospin and temperature.
\section{Quasiparticle-vibration coupling in relativistic framework:
shell structure of superheavy Z=120 isotopes}
\begin{figure}[htb]
\centering
\includegraphics*[width=150mm]{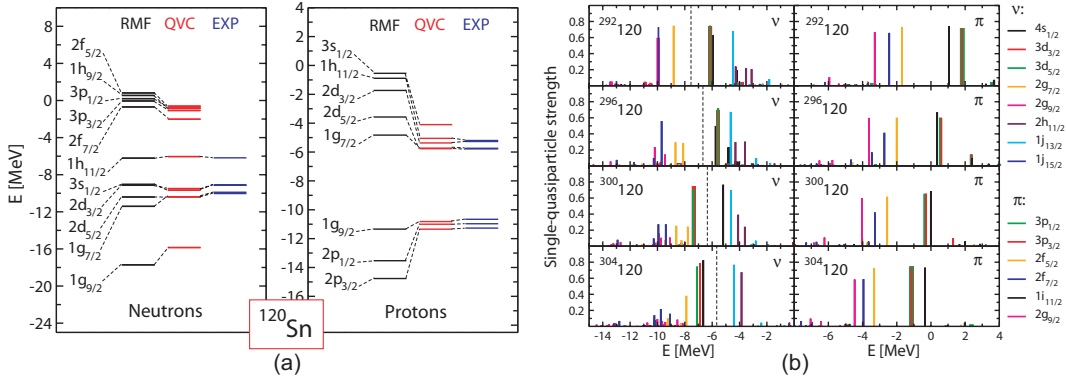}
\caption{(a) Single-quasiparticle spectrum of $^{120}$Sn:
Relativistic mean field (RMF, left column), QVC (center) and
experimental data (right). In the 'QVC' and 'EXP' cases only the
dominant levels are shown. (b) Single-quasiparticle strength
distribution for the orbits around the Fermi surfaces in the neutron
(left panels) and proton (right panels) subsystems of the Z=120
isotopes calculated in the relativistic quasiparticle-vibration
coupling model. The dashed lines indicate the neutron chemical
potentials.} \label{sps}
\end{figure}

We show how the shell structure of open-shell nuclei can be
described in a fully self-consistent extension of the covariant
energy density functional theory. The approach implies
quasiparticle-vibration coupling (QVC) in the relativistic framework
being an extension of the Ref. \cite{LR.06} for superfluid systems
\cite{L.12}. Medium-mass and heavy nuclei represent Fermi-systems
where single-particle and vibrational degrees of freedom are
strongly coupled. Collective vibrations lead to shape oscillations
of the mean nuclear potential and, therefore, modify the
single-particle motion. As a result, single mean-field states split
into levels occupied with fractional probabilities which correspond
to spectroscopic factors of these fragments. The Dyson equation is
formulated in the doubled quasiparticle space of Dirac spinors and
solved numerically for nucleonic propagators in tin isotopes which
represent the reference case: the obtained energies of the
single-quasiparticle levels and their spectroscopic amplitudes are
in excellent agreement with data, see Fig. \ref{sps}(a).
Because of high universality of the approach it can be applied to
nuclei at the limits of their existence with respect to their proton
and neutron numbers, for instance, to superheavy nuclei. Selected
results on the single-quasiparticle strength distributions in the
neutron and the proton subsystems of the Z = 120 isotopic chain are
displayed in Fig. \ref{sps}(b). One can see the evolution of these
distributions with an increase of the neutron number from N = 172 to
N = 184. The shell gap in the proton subsystems of the considered
nuclei diminishes only little when the neutron number increases, so
that the proton number Z = 120 remains a rather stable shell closure
while the detailed structure of the proton levels shows some
rearrangements induced by the neutron addition. In the neutron
subsystems both pairing and QVC mechanisms are active and show a
very delicate interplay: pairing correlations tend to increase the
shell gap while the QVC alone tends to decrease it and at the same
time causes the fragmentation of the states in the middle of the
shell. As a result, in the presence of both mechanisms the gap in
the neutron subsystem remains almost steady while the newly occupied
levels jump down across the gap when the neutrons are added. Thus,
in contrast to the pure mean field studies \cite{ZMZ.05}, no sharp
neutron numbers appear as the candidates for the spherical shell
closures in this region: the shell closures are delocalized.

\section{Spin-isospin response of neutron-rich nuclei}

\begin{figure}[htb]
\centering
\includegraphics*[width=150mm]{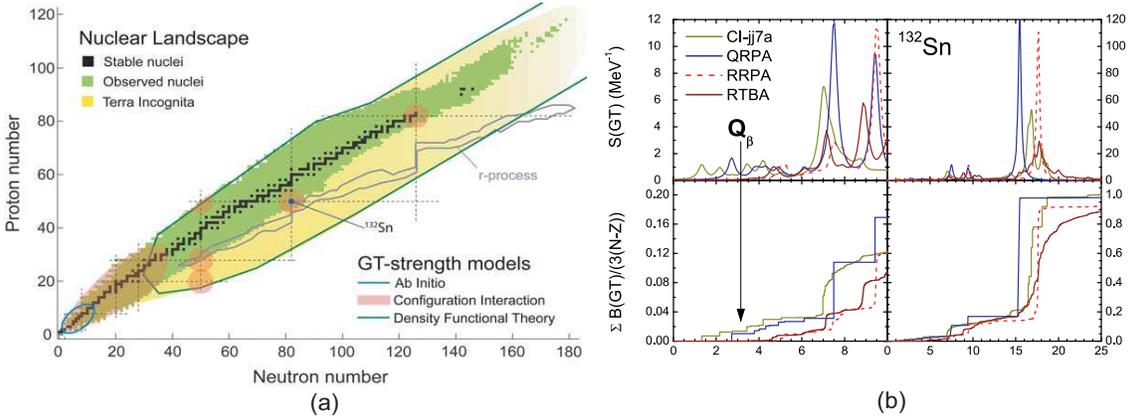}
\caption{(a) Chart of the nuclei \cite{Bertsch} representing stable
nuclei and nuclei found in nature (black), those produced and
investigated in the laboratory (green) and theoretical limits of
bound nuclei (yellow). The domain of ab-initio models is the
lightest nuclei (blue outline), the configuration interaction
approach is applicable in the pink areas and density functional
theory covers the region outlined in green. (b) The Gamow-Teller
strength distribution in neutron-rich $^{132}$Sn calculated within
the various theories (see text for details): low-energy and total
strength distributions (upper panels) and their cumulative sums
(lower panels), respectively.} \label{sir}
\end{figure}

Although last decade the three major concepts in low-energy nuclear
theory (i) ab-initio approaches, (ii) configuration interaction
models (known also as shell-models) and (iii) density functional
theory (DFT) have advanced considerably, they still have to be
further developed to meet the requirements demanded by contemporary
nuclear experiment and astrophysics. Furthermore, each of these
concepts has principal limitations of their applicability in the
nuclear physics domain. Fig. \ref{sir}(a) shows an image of the
nuclear chart taken from Ref. \cite{Bertsch}. The areas of the
nuclear landscape which can be described by each of the three
theoretical concepts are outlined (here we focus on the spin-isospin
nuclear properties, e.g., Gamow-Teller response).

The sectors of the nuclear landscape where the applicability areas
of the different models overlap are of particular interest because
within these sectors the models can constrain each other. Ab-initio
models can replace the phenomenological input which is traditionally
used in the shell-model with the microscopic effective interaction
computed from the first principles. In turn, the shell model with
its very advanced configuration interaction concept can guide the
DFT-based developments beyond its standard random phase
approximation. Thus, in contrast to considering different models as
independently developing alternatives, we rather admit their
complementarity which can be used for their further advancements.

In  Fig. \ref{sir}(b) we present studies of the Gamow-Teller
resonance (GTR) in $^{132}$Sn within the three theoretical concepts:
(i) non-relativistic quasiparticle random phase approximation (QRPA)
with realistic G-matrix interaction \cite{STF88}, (ii) covariant
DFT-based relativistic random phase approximation (RRPA) and its
extension to particle-hole$\otimes$phonon configurations called
relativistic time blocking approximation (RTBA) \cite{MLVR.12}, and
(iii) shell-model (SM) with the configuration interaction CI-jj7a
truncated by the Tamm-Dancoff proton-neutron phonon coupled to
particle-hole core vibrations \cite{hb13}. The gross and fine
features of the GTR obtained within these models are compared, the
advantages and drawbacks of the considered models are discussed.
Based on such comparative studies, future directions are outlined
for each of the above mentioned microscopic models \cite{F.13}.
Constraints on the many-body coupling schemes and underlying
interactions from measurements at the future rare isotope facilities
are anticipated.

\section{Finite-temperature effects on low-energy nuclear response}

Excitation energy is another characteristic of excited nuclei which
imposes limitations of their existence and plays a very important
role in astrophysical modeling. We consider the finite-temperature
effect on the low-energy nuclear response known as upbend
phenomenon, which was first reported in Ref. \cite{V.04}, later
observed systematically in the $\gamma$-ray strength functions below
neutron threshold of various light and medium-mass nuclei and probed
by different experimental techniques \cite{W.12}. Studies of Ref.
\cite{LG.10} have revealed that this phenomenon, occurring in
various astrophysical sites, can have a significant impact on their
elemental abundances.

\begin{figure}[htb]
%\centering
%\includegraphics*[width=15cm]{fig1.eps}
\includegraphics[height=.4\textheight]{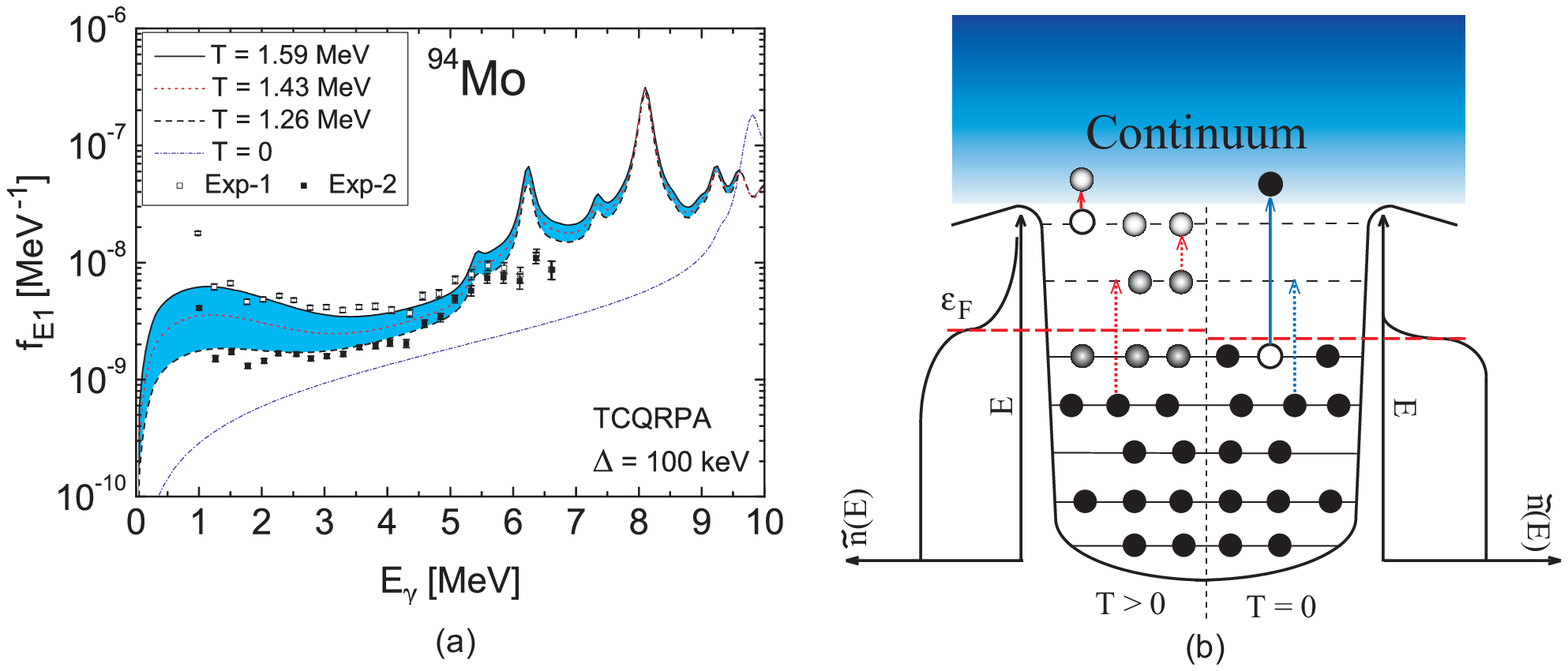}
%\label{94mo}
\end{figure}
\vspace{-30mm}
\begin{figure}[htb]
\caption{(a) The E1 $\gamma$-strength for the thermally excited
state of $^{94}$Mo near the neutron separation energy (blue band),
compared to the strength for the ground state (dash-dotted) and to
Oslo data. (b) Schematic picture of the lowest-energy
single-quasiparticle transitions from the thermally unblocked states
with effective occupation probabilities ${\tilde n}(E)$ to the
continuum.} \label{94mo}
\end{figure}
We propose a microscopic approach for the radiative strength
function which is based on the statistical description of an excited
compound nucleus. The thermal mean field describes the nuclear
excited states of the compound type very reasonably and, at the same
time, it is simple enough to allow a straightforward generalization
of very complicated microscopic approaches to nuclear response in
terms of finite temperature corresponding to the nuclear excitation
energy. To describe transitions from a thermally excited state, in
the first approximation we employ the finite-temperature continuum
QRPA developed in \cite{LKT.03}. The two-quasiparticle propagator in
nuclear medium is calculated in terms of the Matsubara temperature
Green functions in the coordinate space. The continuum part of this
propagator is responsible for transitions from the thermally
unblocked discrete spectrum states to the continuum. The radiative
dipole strength function $f_{E1}$ is determined from the propagator
in the standard way. Fig. \ref{94mo}(a) displays the
$\gamma$-strength in $^{94}$Mo at the excitation energy around its
neutron separation energy, that represents the case of radiative
thermal neutron capture, compared to $\gamma$-strength in the ground
state. The origin of the $\gamma$-strength upbend due to the
transitions to the continuum is illustrated schematically in Fig.
\ref{94mo}(b). The upbend appears as a typical feature of the
$\gamma$-strength in medium mass nuclei while in heavy nuclei the
strength is flat at $E_{\gamma}\to 0$ \cite{LB.13}. The obtained
results have an important consequence for astrophysics, namely for
the approaches to r-process nucleosynthesis: as shown in Ref.
\cite{LG.10}, the low-energy upbend in the $\gamma$-strength can
give rise to a considerable enhancement of the neutron capture rates
in neutron-rich nuclei and, consequently, influences the global
abundance distribution.

Based on the obtained results, we expect further advancements of the
theoretical approaches discussed in this contribution. The proposed
developments on many-body coupling schemes and underlying
interactions will need constraints from data on nuclei away from the
valley of stability. Such data will be obtained in experiments
performed at existing, and vastly enhanced capabilities presented by
future rare isotope facilities.

%\begin{figure}
%  \includegraphics[height=.3\textheight]{golfer}
%  \caption{Picture to fixed height}
%\end{figure}

\bibliographystyle{aipproc}   % if natbib is available

%\bibliography{sample}

\begin{thebibliography}{9}
%
\bibitem{LR.06} E. Litvinova and P. Ring, Phys. Rev. C {\bf 73}, 044328
(2006).
%
\bibitem{L.12} E. Litvinova, Phys. Rev. C 85, 021303(R) (2012).
%
\bibitem{ZMZ.05} W. Zhang et al., Nucl. Phys. A {\bf 753}, 106
(2005).
%
%
%
\bibitem{Bertsch} G.F. Bertsch, J. Phys.: Conf. Ser. {\bf 78},  012005 (2007).
%
\bibitem{STF88} J.~Suhonen, T.~Taigel and A.~Faessler,
Nucl.\ Phys.\ {\bf A486}, 91 (1988).
%
\bibitem{MLVR.12} T. Marketin, E. Litvinova, D. Vretenar, and P. Ring, Phys. Lett. B {\bf 706}, 477 (2012).
%
\bibitem{hb13} M. Horoi and B. A. Brown, Phys. Rev. Lett. {\bf 110}, 222502 (2013).
%
\bibitem{F.13} D.-L. Fang, T. Marketin, B.A. Brown, E. Litvinova,
and R.G.T. Zegers, in preparation.
%
\bibitem{V.04} A. Voinov {\it et al.}, Phys. Rev. Lett. 93, 142504.
%
\bibitem{W.12} M. Wiedeking {\it et al.}, Phys. Rev. Lett. 108,
162503.
%
\bibitem{LG.10} A.C. Larsen and S. Gorieli, Phys. Rev. C 82, 014318
(2010).
%
\bibitem{LKT.03} E.V. Litvinova, S.P. Kamerdzhiev, and V.I. Tselyaev, Phys. Atomic Nuclei 66, 558 (2003).
%
\bibitem{LB.13} E. Litvinova and N. Belov, arXiv:1302.4478.
%
%
\end{thebibliography}

%  \typeout{******************************************}
%  \typeout{** Please run "bibtex \jobname" to optain}
%  \typeout{** the bibliography and then re-run LaTeX}
%  \typeout{** twice to fix the references!}
%  \typeout{******************************************}

\end{document}